
%
\tolerance=10000
\magnification=\magstephalf

\def\J{\hbox{$\bf J$}}

\def\I{\hbox{${\bf I}$}}
\def\H{\hbox{$\bf H$}}
\def\ni{\hbox{$n_i$}}

\def\ri{\hbox{$r_i$}}
\def\rj{\hbox{$r_j$}}
\def\mi{\hbox{$m_i$}}
\def\mj{\hbox{$m_j$}}
\def\Ti{\hbox{$T_i$}}
\def\Tj{\hbox{$T_j$}}

\def\Sig1N{\hbox{$${\sum_{j=1}^N}$$}}
\def\Sig1i{\hbox{$${\sum_{j=1}^i}$$}}

\def\Tij1{\hbox{$(\Ti + \Tj)^{1/2}$}}
\def\rij2{\hbox{$(\ri + \rj)^2$}}

\def\u2{\hbox{$u^2$}}

\def\os2{\hbox{${1\over \sqrt{2}}$}}

\def\mi{\hbox{$m_i$}}
\def\mj{\hbox{$m_j$}}

\def\mo{\hbox{$m_o$}}

\def\ml{\hbox{$m_l$}}

\parindent0pt
\baselineskip=.6truecm
\topskip=.8truecm
\null
\centerline{MASS SPECTRUM AND VELOCITY DISPERSIONS}
\vskip .3cm
\centerline{DURING PLANETESIMAL ACCUMULATION}
\vskip .3cm
\centerline{(II - FRAGMENTATION )}
\vskip 4.5truecm

\line{\hskip 2cm P. Barge\hfill}
\line{\hskip 5cm Ecole Normale Sup\'erieure de Lyon\hfil}
\line{\hskip 5cm 46 All\'ee d'Italie \hfill}
\line{\hskip 5cm 69364 - Lyon - C\'edex 07 - France \hfill}
\vskip .2cm
\line{\hskip 5cm Laboratoire d'Astronomie Spatiale\hfill}
\line{\hskip 5cm  et Observatoire de Marseille\hfill}
\line{\hskip 5cm B.P. 8 \hfill}
\line{\hskip 5cm 13376 - Marseille - C\'edex 12 France\hfill}
\vskip 1truecm
\line{\hskip 2cm R. Pellat \hfill}
\line{\hskip 5cm Groupe de Physique Th\'eorique\hfill}
\line{\hskip 5cm de l'Ecole Polytechnique \hfill}
\line{\hskip 5cm F-91128 Palaiseau - France \hfill}
\vskip 3.5truecm

pages : 30  \hfill\par
\vskip 0.4truecm
figures : 8 \hfill\par
\vskip .4truecm
Tables : 0  \hfill\par
\vfill\eject

\topskip=4truecm
\line{Running Title :  Protoplanetary Cloud\hfill}
\vskip .5cm
\line{Key words :  Collisions - Plasmas Physics Protoplanetary Cloud\hfill}
\vskip 3truecm
\line{Address for proofs : \hfill}
\vskip .5cm
{}~~~~~~~~~~~~~~~~~~~~~~P. Barge
\hfill\par
{}~~~~~~~~~~~~~~~~~~~~~~Laboratoire d'Astronomie Spatiale \hfill\par
{}~~~~~~~~~~~~~~~~~~~~~~B.P. 8
\hfill\par
{}~~~~~~~~~~~~~~~~~~~~~~13376 - Marseille - C\'edex 12 - France \hfill\par
{}~~~~~~~~~~~~~~~~~~~~~~Tel :  91 05 59 00
\hfill\par
{}~~~~~~~~~~~~~~~~~~~~~~Electronic Mail Address :
\hfill\par
{}~~~~~~~~~~~~~~~~~~~~~~EARN (BITNET) BARGE@FRLASM51 \hfill\par
{}~~~~~~~~~~~~~~~~~~~~~~SPAN (DECNET) LASM0B::BARGE
\hfill\par
\vfill\eject

\baselineskip=.5truecm
\topskip = 2truecm

\line {ABSTRACT\hfill}

\vskip .4cm
The increase of the velocity dispersions which occurs during the growth of
the planetesimals (Barge and Pellat, 1991) strongly suggested that
fragmentation could come at work in the final stages of the accumulation
scenario.
\hfil\par
Fragmentation has been modeled assuming that mass and energy are redistributed
in a  simple way between the various fragments; it is found to play a role
whose importance is greater than previously believed and depends on the average
characteristic size of the primordial planetesimals but not on their mass
distribution. In the small size range the mass spectrum is strongly modified
with the formation of a small bodies tail but, on the other hand, the growth of
the most massive objects is significantly slowed down. With the recent
fragmentation model of Housen et al (1991) it is found that 10 kilometer-sized
objects can grow into planetary embryos (size of the order of 1000 km) in a
time-scale of the order of $10^5$ or $10^7$ years (depending on the
fragmentation properties of the planetesimals) whereas 1 kilometer-sized
objects cannot on a reasonable time-scale. The formation of planetary embryos
much more massive than the rest of the swarm is delayed; nevertheless the
time-scale problem for the growth of the outer planets is not revived. The
elasticity of the collisions is found to play a negligible role in the
investigated size range.
\hfil
\vfil\eject

\baselineskip .5cm

\topskip = .5truecm

\line{I) Introduction \hfill}

\vskip .3cm

Catastrophic collisions of the planetesimals are often considered in the
literature as plausible, if not unavoidable, events which could have played a
role during the accumulation of the planets (Safronov, 1969; Greenberg et al,
1978). Indeed it is likely that the destruction of bodies with size larger than
a kilometer could affect the mass spectrum and the velocity dispersion in the
whole swarm of planetesimals through mass and energy transfers from the large
scales to the small scales. The occurrence of catastrophic collisions has also
been suggested to account for a number of observations in the present solar
system (Fujiwara, 1982; Davis et al , 1985); they could explain, for example,
the existence of some dynamical families in the asteroid belt (Davis et al,
1979) and the irregular shape of Hyperion around Saturn (Farinella et al,
1983).
\hfil\par
The collisional destruction of the planetesimals is poorly understood due to
the complexity of the physical mechanisms coming into play and also to the lack
of knowledge about the material and the internal structure of these bodies. In
spite of this, as the resulting mass and energy redistribution could
modify the accumulation scenario, fragmentation has to be
modeled, even if in a crude way.
\hfil
\vskip .2cm
The evolution of asteroidal bodies under collisional fragmentation only was the
first problem to be investigated (Piotrowski, 1953; Dohnanyi, 1969; Hellyer,
1970). Using a geometrical cross-section, a constant impact velocity and
assuming further the fragment mass distribution follows an inverse power law
these authors found a steady state solution of the fragmentation equation:
except for the largest bodies, the mass distribution is a decreasing power law
with exponent 0.8. This result seems in agreement with the size distribution of
the asteroids and meteorites and the statistics on the lunar craters.
\hfil
\par
\vskip .2cm
The accumulation of the planetesimals proceeds from a long term evolution and
the assumption of a constant impact velocity is clearly irrelevant; in fact,
the occurrence of fragmentation depends on the velocity dispersions of the
colliding bodies and fragmentation itself modifies the velocity dispersions
through mass and energy redistribution. This self-consistent aspect of the
problem (the same as that encountered in the growth of indestructible bodies)
must be taken into account in the evolution of the planetesimals.
\hfil
\vskip .2cm
A model of the planetesimal accumulation which accounts both for accretion and
fragmentation has been performed first by the "Moscow School" (Zvyagina et al,
1974; Pechernikova, 1975; Pechernikova et al, 1976). Starting from a power law
fragment mass distribution (with exponent 0.8) and assuming
the velocity dispersions increase approximately as the escape velocity of the
largest body, these authors solve numerically a generalized coagulation
equation and found that the mass distribution evolves differently following the
size range: at the small scales the mass spectrum can be fitted by a decreasing
power law which tends slowly toward an asymptotic solution, whereas at the
large scales the time dependence is stronger but no asymptotic behaviour can be
established. Unfortunately, estimating the velocity dispersions, these authors
omitted to account for the effect of the dynamical friction.
\hfil
\vskip .2cm
Later Greenberg et al (1978, hereafter G78) and then Wetherill and Stewart
(1989, hereafter WS89) considered also the occurrence of fragmentation during
planetesimal growth. Both take account of the dynamical friction: Greenberg et
al in a crude way with their pioneer numerical simulation, Wetherill and
Stewart through a formalism developed earlier (Hornung et al, 1985) but used
outside its range of validity. Both alike found a velocity dispersion of the
largest bodies much smaller than derived with a more elaborate three-body model
(Barge and Pellat, 1991) and, by the way, underestimated the actual importance
of the fragmentation.
\hfil
\vskip .1cm
Beaug\'e and Aarseth (1990) have also accounted for the occurrence of
fragmentation in their N-body simulations of the last stage of the planetary
formation. Starting from 200 identical lunar-sized bodies they found that large
embryos are formed very early in the evolution and that the gravitational
perturbations tend to increase the mean eccentricity till fragmentation comes
into play; then creation of new embryos stops, those already formed continue to
grow while fragments decrease in size. However it is to be noted that their
results are not directly comparable with the present ones, on the one hand
because their computations are only two dimensional and on the other because
they start from lunar-sized bodies which are much stronger against
fragmentation than the kilometer-sized first-born planetesimals we started
from.
\hfil
\vskip .1cm
In paper I we studied the simultaneous evolution of mass and velocity
dispersions of accreting indestructible planetesimals: the largest bodies were
found to grow very fast with an increasing velocity dispersion. The evolution
of the largest bodies, in spite of appearances, contrasted with that of
Safronov (1969) who neglected completely the dynamical friction and found much
longer time scale but it contrasted also with that of Wetherill and Stewart
WS89 who claimed that the velocity dispersions of the largest bodies remain
always very weak (weaker than our values by more than an order of magnitude).
This increase of the velocity dispersions suggested that fragmentation might
play a role in the final stages of the planetesimal accumulation.
\hfil
\vskip .1cm
In this paper we will devise a simple model of the planetesimals fragmentation,
based on laboratory results and on a recent model for the collisional
disruption of the asteroids, in order to draw nearer to the final stages of the
accumulation scenario. Obviously many features of the destruction mechanisms
will be neglected as being too detailed for our degree of extrapolation but on
the other hand the assumption of purely inelastic collisions will be avoided.
This work attempts a self consistent approach of the planetesimal accumulation
in which fragmentation is included in the kinetic description of the
planetesimal evolution. The accumulation proceeds by a competition between
accretion and fragmentation; the evolution of the velocity dispersions account
for collisions and encounters as in our previous papers, but also for
fragmentation through a simple energy variation rate.
\hfil
\vskip .2cm
First we reported in section II the results obtained by improving the
accretion-code developed in paper I. Section III is devoted to an overview of
the fragmentation process and to the determination of the energy thresholds for
simple disruption or complete fragmentation of the planetesimals. Section IV
illustrates the occurrence of the various possible outcomes of the planetesimal
collisions. The mass and energy budget necessary to model fragmentation is
performed in section V. The complete set of the generalized evolution equations
is given in section VI. The results of the integrations of the evolution
equations are reported in section VII for various initial conditions. Finally
section VIII is devoted to the conclusions and to the discussion of the
results.
\hfil

\vskip .7cm
\line{II) The growth of indestructible planetesimals\hfill}

\vskip .3cm
Let us first take a second look at the accretion calculations performed in
paper  I. The growth of the planetesimals is described in a Lagrangian approach
 of the coagulation equation, modified in the course of the integration as to
account for mass dispersion. From the Lagrangian point of view the evolution
of the mass distribution is described with moving batches and the collisional
growth process offers two different aspects:
 \hfil
\vskip .1cm
- large bodies sweep up smaller ones in the same way as vapor condensates onto
liquid droplets; this mechanism is responsible for batch motion in concerted
growth.
 \hfil
\vskip .1cm
- nearly identical bodies coalesce and transfer the newly created entities to
neighbouring batches in the same way as liquid droplets coalesce into larger
ones; this mechanism is responsible for mass dispersion among batches.
 \hfil
\vskip .1cm
Both aspects were taken into account in our accretion code except at the
largest size end, since, if total number of batches is kept constant, it is
impossible to transfer the largest bodies to larger batches.
\hfil
\par
Nevertheless this incomplete description of the evolution is of no consequence
if mass dispersion is negligible (as occurs for example when the initial mass
distribution is the Safronov exponential we started from in paper I); on the
other hand it is a serious limitation when starting from less steeper mass
distribution as simple power laws.
\hfil
\par
The code has been improved by opening new batches at masses larger than the
largest bodies; initially empty they fill up gradually as a function of time in
the same procedure as in paper I. This improvement was made at the beginning of
the evolution when mass dispersion have important consequences (as is the case
when the population of the largest bodies is very large).
\hfil
\par
The other problem encountered in our re-examination of paper I is related to
the fact that the thickness of the swarm of planetesimals is inversely
proportional to the velocity dispersion. In paper I the variation of the volume
of the cloud was taken into account in an incorrect way. In fact, the
coagulation coefficient is proportional to the gravitational focusing factor;
this has been put right in all subsequent calculations.
\hfil
\vskip .1cm
Then, with these two improvements of the code, new computations were performed
with 12 batches and starting from the same mass spectrum as in paper I; that
is:
\hfil
\vskip .1cm
(i) a Safronov exponential (the same as used by WS89) in which the bodies have
a size range between 4.89 km and 14.8 km;
\hfil
\vskip .1cm
(ii) decreasing power laws in which the size of the bodies ranges from 0.84 km
to 1.26 km.
\hfil
\vskip .1cm
In both cases the calculations were carried out until a single body remains in
the largest batch; the results we found have been plotted on figures 1, 2 and 3
and show that this massive body detaches clearly from the rest of the mass
spectrum.
\hfil
\vskip .2cm
Starting from an exponential, the evolution of the planetesimals is quite
similar to that obtained in paper I; it is found that the largest bodies behave
differently from the rest of the mass spectrum and have two main
characteristics:
\hfil
\vskip .1cm
(i) their velocity dispersion first decreases below the Hill velocity (due to
the dynamical friction) and then increases above this limit (due to the viscous
stirring),
\hfil
\vskip .1cm
(ii) their growth, very rapid at the beginning (when the velocity dispersion is
very low), tends to slow down (in accordance with the thickening of the swarm).
\hfil
\vskip .1cm
At the end of the evolution, after $3.10^5~yrs$, a single massive embryo
succeeded in capturing approximately 17$\%$ of the total mass and its orbit
have an eccentricity of the order of $4.10^{-3}$. It must be emphasized that
this value of the eccentricity is much larger than found by WS89 or G78.
\hfil
\vskip .2cm
Starting from power laws, the evolution of the mass distribution is different
from that obtained in paper I. Two different cases have been distinguished
following the mass of the swarm is initially dominated by the large bodies
(exponent equal to 0) or by the small ones (exponent equal to -2) but, in both
cases, the evolution is approximately the same:
\hfil
\vskip .1cm
- the mass distribution relaxes rapidly into a very steep function (even
steeper than the Safronov exponential) due to the kinetics of the coagulation
process itself. After this transient stage the evolution is similar to that
obtained when starting from the Safronov exponential.
\hfil
\vskip .1cm
- the velocity dispersions, after a quick relaxation toward a near
equipartition of the energy, increase as a function of time; as above the
largest bodies can reach values greater than the Hill velocity at the end of
the evolution.
\hfil
\vskip .2cm
After $2. 10^5~yrs$ a single embryo contains some $16\%$ of the total mass and
its orbit have an eccentricity of the order of 0.012 . The difference we find
between these results and those of paper I results mainly from the account of
the mass dispersion among the largest bodies in the first part of the
integration (a possibility which was excluded in our previous computations).
\hfil
\vskip .1cm
To summarize, a transient behaviour appears in the beginning of the evolution
of the planetesimals: the mass distribution relaxes into a very steep function
which gathers most of the total mass into the smallest bodies and,
consequently, the distribution of the velocity dispersions relaxes according to
a near equipartition of the energy; this behaviour indicates what the initial
state of the swarm, compatible with the kinetics under encounters and inelastic
collisions, is. Then the evolution is the following:
\hfil
\vskip .1cm
(i) first the weak velocity dispersions of the largest bodies produces strong
gravitational focusing and, so, favours start of "runaway";
(ii) as time
elapses, most of the total mass tends to shift from the small scales to the
large ones so that gravitational stirring gradually tends to overcome
dynamical friction (see paper I).
As a result the tendency to "runaway" weakens and the
mass tends to stream in a more progressive way to the larger scales.
\hfil

\vskip .7cm
\line{III) The destruction of colliding planetesimals\hfill}

\vskip .3cm
Now we will set out some general considerations about the collisional
disruption mechanism. To begin with, it is obvious that colliding bodies
shatter if the energy released at the impact is strong enough to overcome
material's solid-state cohesion and that the number and the size of the
resulting pieces will depend on the initial kinetic energy.
\hfil
\par
Further informations about the disruption mechanisms come from laboratory
experiments and statistics of the asteroid observations.
\hfil
\vskip .2cm
$\bullet$ The laboratory experiments, performed with small projectiles fired
against plane surfaces (Gault et al, 1963; 1969) or against decimetric basalt
or mortar targets (Fujiwara et al, 1977; Bianchi et al, 1984; Davis and Ryan,
1990; Ryan et al, 1991; Housen and Schmidt, 1991), bring some general results
frequently extrapolated to the much larger scales. In the case of the
planetesimals the most useful ones to modeling are:
\hfil\par
\vskip .2cm
(i) the mass distribution of the fragments obeys approximately a decreasing
power law with an exponent ranging from 2/3 for barely catastrophic collisions
to 1 for supercatastrophic destructions,
\hfil
\vskip .1cm
(ii) the velocity distribution of the fragments is roughly fitted by a
decreasing power-law with an index equal to -1/6 (Nakamura and Fujiwara, 1991),
\hfil
\vskip .1cm
(iii) in the case of core type destructions the relation between impact energy
and size of the largest fragments can be fitted by a power law with an exponent
equal to 1.24,
\hfil
\vskip .1cm
(iv) the actual shape of the target seems unimportant as long as the aspect
ratio is of the order unity.
\hfil
\vskip .2cm
A qualitative physical interpretation of disruptions by high velocity impacts
was given by Gault and Wedekind (1969), by Fujiwara (1980) and by Housen and
Holsapple (1990). After the impact a compressive wave expands radially into the
target and decays very quickly; then, if the body is smaller than a certain
size, this wave reflects off the free surface of the target as a complex system
of shear and tension waves which, due to the relatively weak tensile strength
of rocks, results in multiple spallation and in a destruction of the solid
body (see also Fujiwara, 1982 and Fujiwara and Tsukamoto, 1980). This fracture
process occurs as a result of the growth and coalescence of pre-existing cracks
whose activation depends on the rate at which the material is loaded (Housen
and Holsapple, 1990).
\hfil\par
However in the laboratory experiments the impact energy never exceeded some
$10^3J$ which is much smaller than the energy necessary for the break up of
asteroid like bodies which ranges rather from $10^{15}J$ to $10^{20}J $.
\hfil\par
\vskip .2cm

$\bullet$ The observation of asteroid families in the present solar system
brings also some useful informations about the very large scale collisions.
Indeed the mass distribution of asteroids with strongly correlated orbits has
been interpreted by most authors (Anders, 1965; Dohnanyi, 1969; Chapman and
Davis, 1975; Ip, 1979) as being produced by a catastrophic collisional
evolution. In some cases it has been possible to reconstruct the parent bodies,
to give a rough estimate of the impact energy (Fujiwara, 1982; Davis et al,
1985) and even to speculate about the internal structure of the parents.
\hfil
\par
The mass distribution in the asteroid families have a power law exponent in
agreement with the high velocity experiments (Fujiwara, 1980) and moreover the
mean relative velocities between the family members are not incompatible with
the ejection velocities of the laboratory fragments (Zappala et al; 1984). So
extrapolation of experimental datas to the much larger scales, although rather
tricky seems reasonable.
\hfil
\vskip .2cm
Fujiwara (1980) extrapolated over many order of magnitudes the relation he
found between critical size for disruption and impact energy (a power law with
an index $\zeta$ between 0.36 and 0.44); then he suggested that the
experimental results could be applicable to gravity free bodies even with size
as large as $100~km$. It must be noticed that, if $\zeta$ is larger than $1/3$,
the impact strength decreases with the body size; this idea has been used by
Farinella et al (1982) to conclude that asteroids could be intrinsically weak
as a simple consequence of their large size and corresponds to the assumption
that the fragmentation energy is proportional rather to the total surface area
of the produced fragments than to their volume. At the time this idea
contrasted with the classical assumption of an impact strength independent of
the body size (Greenberg et al, 1978; Davis et al, 1979; Fujiwara, 1982).
\hfil
\par
However very recently Housen and Holsapple (1990) constructed a model,
based on a variety of experimental and theoretical evidences, in which the
disruption strength depends not only on target size and impact energy but
also on strain-rate and impact velocity; this model leads again to the
conclusion that large bodies are weaker than small ones, suggesting that
asteroids should fracture at much lower specific energy than small-scale
experiments would suggest.
\hfil
\par
\vskip .2cm

In the case of the planetesimals it must be emphasized that no reliable data
exist about the fragmentation or even the composition of these bodies. They
could be constituted from some primitive carbonaceous chondritic material
resulting from the aggregation of solar system rocky condensates whose
mechanical properties would be comparable to the terrestrial basalt. Moreover
the planetesimals, as a number of asteroids in the present solar system, could
be coated with a thick regolith layer or again strongly fractured by the
repeated collisions; so the propagation of the compressive waves into their
interiors would be very different from that occurring in the experimental
cases. Nevertheless some idea becomes to emerge about the large scale
collisional fragmentation. In recent year Housen et al (1991) conducted new
fragmentation experiments simulating these large scale catastrophic events
(targets with size larger than some 100 km) by applying an appropriate
overpressure onto small decimetric targets; they found that the specific
energies for fragmentation are of the same order of magnitude as those
estimated from the observations of the Themis, Eos and Koronis asteroids
families.
\hfil
\par
It is also to be noted that during the destruction of bodies with size larger
than $100~km$, the gravitational attraction between the resulting pieces must
be also taken into account as being able to hinder seriously the dispersal of
the fragments. Thus, after a break-up event, it is necessary to distinguish
between two cases: the simple disruption of the body cohesion and the complete
"fragmentation", that is a shattering of the body followed by a dispersion of
the resulting pieces as independent fragments.
\hfill
\vskip .5cm

\line{1) Disruption threshold\hfill}

\vskip .2cm
Following Housen et al (1991) the energy threshold for the disruption of a
planetesimal $m_i$ can be written:

$$ Q_D(i) = \left[Q_o \left({\mi \over m_o}\right)^{a}   + G_o \left({\mi
            \over m_o}\right)^{b}\right] \left( {\Delta V}\over {V_A}
                                         \right)^{0.35} ~~.  \eqno (3.1) $$

\vskip .1cm
where $a = 0.92$, $b = 1.55$, $\Delta V$ is the impact velocity and $V_A$ is
the average relative velocity in the asteroid belt ($\simeq 5~ km s^{-1}$);
$Q_o = 1.5~10^3~m_o^{0.92} $ is scaled from experimental values only and $ G_o
= 7.7~10^{-7}~m_o^{1.55} $ is scaled both from the new experimental values of
Housen et al (1991) and from the disruption specific energies estimated by
Fujiwara (1982) from a reconstruction of parents bodies for the Themis, Eos
and Koronis Hirayama families (for a kilometer-sized bodies $Q_o \simeq
2.1~10^{15} J$ and $G_o \simeq 2.3~10^{14} J$). The first term in (3.1)
corresponds to the energy necessary for the break up when the body is
considered as a gravity free solid whereas the second one, introduced first by
Davis et al (1985), is the energy necessary to overcome material strengthening
due to the gravitational compressive load. Indeed, due to the variation of the
pressure with depth, the interiors are more difficult to disrupt than the near
surface and the intensity of the tension waves must exceed the material
strength plus the compressive load of the upper layers.
\hfil
\vskip .2cm
Obviously in the case of the planetesimals the effective strength is highly
uncertain due to the lack of knowledge on their actual composition and internal
structure. They could be brittle bodies made either of weakly bounded regolith
(G78) or of deeply cracked material but, on the contrary, they could be much
more stronger if constituted by a thick regolith layer upon a solid core
(Hartmann, 1978). At any rate the energy threshold (equation (3.1)) deduced
from the above mentioned considerations is, at present, the most detailed
expression available to describe planetesimal evolution; it
will be used through the rest of the paper.
\hfil
\vskip .2cm
Now, let us check the two simplifying assumptions we have made as to model
a catastrophic event (that is when the impact energy is large enough to
shatter the target) in a tractable way:
\hfil
\vskip .1cm
 - both impacting bodies shatter in various fragments;
\hfil\par
this assumption is appropriate in the case of identical bodies, but must be
discussed in the opposite case. If the projectile is shattered and the target
is not, two cases have to be distinguished: (i) if the target mass is
large enough to capture the various pieces, the collision results in an
accretion of the projectile onto the target (this happened, for example, in the
meteor crater event); (ii) if the target cannot reaccumulate a large fraction
of the projectile mass the assumption is weak and tends to underestimate the
importance of the fragmentation.
\hfil
\vskip .1cm
   - the disruption occurs with the same crushing law for the two impacting
bodies;
\hfil\vskip .1cm
 the destruction of the two bodies is assumed to produce two different cores
and two species of fragments regardless of a "mixing" of the ejectas; this is
a convenient but a rough assumption.
\hfill
\vskip .5cm

\line {2) Fragmentation threshold \hfil}

\vskip .2cm
If the two impacting planetesimals are massive enough the self gravity of their
various pieces can be sufficient for fragments not to escape; then collisions
result in accretion with the formation of a single body rather than in
fragmentation. Bodies issuing from the reaccumulation of a large number of
various pieces could exist among the present asteroids and have been evoked
many times in the literature (Davis and Chapman, 1977; Davis et al, 1979;
Farinella et al, 1982) but whether reaccumulation of fragments can be efficient
enough to produce "piles of rubble" is still an open question. On the other
hand the continuous bombardment by the smaller bodies (and the partial
reaccumulation due to the repeated collisions) leads certainly to a regolith
layer which, if thick enough, could modify seriously the efficiency of the
sticking and disruption mechanisms (Hartmann, 1978).
\hfil
\par
The total amount of energy necessary for a disruption followed by an ejection
of independent fragments will be estimated, in the same way as in the work of
Zvyagina et al (1974), with the help of total energy conservation before and
after the destruction. The energy budget for the fragmentation of two bodies
$m_i$ and $m_j$ will be settled in the following way:
\hfil
\vskip .2cm

\line {- before the impact:~~~~$K_o + U_i + U_j$ ~, \hfill }

\vskip .1cm
where $K_o$ is the initial kinetic energy and $U_i = 3Gm_i^2/5R_i$ is
the gravitational potential energy of the body $m_i$;
\hfil
\vskip .2cm
\line {- after the impact:~~~~$K_{1}  + (1 - \chi)K_o + U_F(i,j) $ ~,\hfill}
\vskip .1cm
where $U_F(i,j)$ is the potential energy of the dispersed fragments and $K_1$
their final kinetic energy; $\chi$ is the fraction of the initial kinetic
energy transferred to the various pieces. After Fujiwara the most probable
value of the $\chi$ factor in the case of asteroidal bodies is only some ten
percent; in all subsequent calculations we will take $\chi = 0.1$ .
\hfil
\vskip .2cm

Finally, energy conservation leads to: $K_{1} =\chi K_o + U_i + U_j -
U_F(i,j).$
\hfill
\vskip .2cm

The dispersal of the fragments will be assumed to occur if their final kinetic
energy is greater than zero. This is obviously an approximate condition which
neglects any mass dependence in the fragment velocity (in fact, small pieces
are easier to disperse than large ones); it is satisfied if the impact energy
is strong enough to overcome an amount of energy equal to:

$$ {1\over \chi} (U_F(i,j) - U_i - U_j)~. \eqno (3.2)$$

\vskip .1cm
On the other hand the potential energy of the dispersed fragments will be
estimated from the continuous mass spectrum $n(m)$ through the relation:

$$ U_F(i,j) \simeq  - \int dm~n(m) U(m)~. \eqno (3.3) $$

\vskip .1cm
As discussed above the differential mass distribution of the fragments is
approximately given by the power law:

$$ n(m) = n_o \left({m\over m_o}\right)^{-q}~, \eqno (3.4) $$

with $q = 1+\alpha $ and $\alpha\simeq 2/3$; then we find approximately:

$$ U_F(i,j) \simeq {{U_o}\over 3}
\left({{m_i + m_j}\over m_o}\right)^{5/3}  \eqno (3.5) $$

where ${U_o} = - {3G m_o^2/5 R_o} $ is the gravitational potential of a typical
body (of the order of $10^{13} J$ for a 1 km body). As a result the energy
threshold for the fragmentation of the two bodies $m_i$ and $m_j$ will be
defined as:

$$Q_F(i,j) = Q_D(i) + Q_D(j) - {{U_o}\over\chi}
            \left[({m_i\over m_o})^{5/3} + ({m_j\over m_o})^{5/3} - {1\over 3}
                  ({{m_i+m_j}\over m_o})^{5/3}\right]~. \eqno (3.6) $$

\vskip .1cm
The second part of the r.h.s. of this equation represents the additional
gravitational binding energy to be exceeded for a dispersal of the shattered
bodies into independent fragments.
\hfil
\vskip .2cm
Then, in order to illustrate our model, the specific energies $Q_D^*$ and
$Q_F^*$ have been plotted on figure 4 as a function of the target size
(assuming an average impact velocity $\Delta V = V_A $); the
results of the experimental tests of Housen et al (1991) and the specific
disruption energies estimated by Fujiwara (1982) from the observations of the
Themis, Eos and Koronis Hirayama families have been reported on the same
figure. As noted Housen et al the reason that the family datas lay above the
$Q_D^*$ threshold is that reaccumulation of collisional debris increases the
size of the largest remnant. This is likely why the $Q_F^*$ fragmentation
threshold, which accounts roughly for reaccumulation effects, seems to give
better agreement between the model and the family datas.
\hfil
\par
Obviously the gravitational attraction between projectile and target is also of
importance before impact since it enhances both collisional cross section and
impact velocity; in our model the gravitational focusing is described only in
a two-body formalism (as in our previous papers) but, on the other hand, the
increase of the impact energy is neglected.
\hfil
\vskip .2cm
Now, before to model the way in which mass and energy are redistributed among
fragments, let us first discuss the occurrence of the various collision
outcomes as a function of the impact energy.
\hfil

\vskip .7cm

\line {IV) The outcomes of the planetesimal collisions \hfil}

\vskip .3cm
An alternative to the catastrophic outcomes is obviously the rebound of the two
bodies with a change of momentum and a loss of energy. This case was studied in
our previous paper with the set up of a critical velocity $u_R$ which
determines whether the two bodies stick one another or rebound. The energy
threshold above which rebounds are expected is derived in a
straightforward way from the expression of $u_R$ and writes:

$$ K_R = {(1 - \gamma_o^2)\over\gamma_o^2} {1\over 2} m_{ij}^* T_{esc}
\eqno (4.1) $$

where $m_{ij}^*$ is the reduced mass of the colliding bodies; $\gamma_o$ is the
restitution coefficient and $T_{esc}$ is the square escape velocity.
\hfil
\par
Now in order to discuss more easily the occurrence of the various
outcomes of a planetesimal collision, the disruption threshold $Q_D$ will be
assumed independent of the impact velocity (that is we will take $\Delta V~ =
{}~V_A$ in equation (3.1)).
\hfill

\vskip .5cm

\line {1) The various outcomes \hfil}
\vskip .2cm

Clearly the different outcomes of a planetesimal collision can be distinguished
when the impact energy is compared with the above defined energy thresholds;
nevertheless the discussion will be easier in term of the velocities
we will define the square velocity thresholds:

$$ T_\nu = {{2 E_\nu}\over m_{ij}^*} \eqno (4.2) $$

where $E_\nu$ stands for one of the energies $K_R$, $Q_D$ or $Q_F$ .
\hfil\par
If one of the two bodies is much smaller than the others ($\mj << \mi$) these
expressions reduce to:

$$T_R~\simeq ~{(1 - \gamma_o^2)\over \gamma_o^2} T_{esc} \eqno (4.3)$$

$$T_D~ \simeq ~\left[{Q_o\over m_o}~({m_i\over m_o})^{a-1} +
                     {G_o\over m_o}({m_i\over m_o})^{b-1}\right]
                    ~{m_i\over m_j} \eqno (4.4)$$

$$T_F~ \simeq ~T_D  + {1\over 5} {T_{esc}\over\chi} {m_i\over m_j}
   \eqno (4.5)$$

and remain available in order of magnitude even for identical bodies. In the
following it will be convenient to name target the large body and projectile
the small one.
\hfill
\par
The three velocity thresholds have been plotted on figures 5 as a function of
the projectile to target radius ratio; for a given size of the target, they
delimit distinct regions which correspond to the various outcomes of a
collision.
\hfil
\vskip .2cm

$\bullet$ In the case of the large targets (sizes of the order of $100~km$;
cf figure 5c) four distinct regions appear. In the upper part, the region of
the strong impact velocities, collisions result in a fragmentation whereas in
the lower part they result in a sticking. What happens in the intermediate zone
is two fold:
\hfil
\vskip .2cm
(i) in the region between the disruption and the fragmentation limits the two
bodies are disrupted but reaccumulate in a single body (a "rubble pile"),
\hfil
\vskip .1cm
(ii) in the region between the rebound and the disruption limits the two bodies
can rebound after the impact; however the existence of this region depends
clearly on the restitution coefficient and on the target mass.
\hfil
\vskip .2cm

$\bullet$ In the case of the small targets (figures 5a and 5b) the velocity
thresholds are lower and the conclusions are similar; the rebound region is
wider but on the other hand the reaccumulation region is narrower or inexistent
since gravitational effects are less important.
\hfil
\vskip .2cm

Obviously, in this simple sketch, erosion and partial reaccumulation are
omitted; moreover complete reaccumulation is taken into account only in a
rough way through the definition of the composite energy threshold $Q_F$.
In fact, it is clear that the crushing law itself should account for gravity
induced mass segregation.
\hfil
\vskip .1cm

Now we will mention some peculiarities of our model:
\hfil
\vskip .2cm
  (i) at the frontier between reaccumulation and rebound, self gravity plays a
crucial role and the outcome of a two body collision is uncertain since an
impact could result either in two "cracked bodies" or in a single weakly
agglomerated structure (this unstable situation disappears if the bodies have
sizes smaller than some $10~km$);
\hfil
\vskip .1cm
    (ii) at the frontier between reaccumulation and sticking, collisions result
either in a single body or in a rubble pile;
\hfil
\vskip .1cm
    (iii) at the intersection between the disruption and the rebound
thresholds, collisions result either in rebound or in sticking or again in
reaccumulation; this situation (which occurs if the projectile mass reach a
critical value $m_c$) have some remote similarity with the coexistence in
thermodynamics of the three different phases of a pure element.
\hfil
\vskip .2cm
However we must keep in mind that such unstable situations emerge from a naive
discussion of a simple model and, so, could remain only speculations.
\hfil
\par
Anyway this simple sketch illustrates easily the occurrence of the various
outcomes and the fact that the accretion of two planetesimals results either
from simple sticking or from fragment reaccumulation. On the other hand the
extent of the rebound region depends on a critical value of the mass which,
normalized to the target mass, writes:

$$ {m_c(i,j) \over m_i} = {{\gamma_o^2}\over{1-\gamma_o^2}} {{3 G_o}\over
                          {10 U_o}}~\left({m_i\over m_o}\right)^{{5\over 3} -b}
\left[ 1 + {Q_o\over G_o}~\left({{m_o}\over {m_i}}\right)^{a-b} \right] ~~.
\eqno (4.6) $$

It is wider for stronger elasticities or smaller targets. Figures 5, with a
restitution coefficient equal to 0.2 (a probable value in the case of the
planetesimals, see for example G78) shows that the rebound region can be very
wide for the smallest bodies.
\hfil
\vskip   .5cm

\line{2) The characterization of the various outcomes \hfill}

\vskip .2cm
Now, in order to proceed further, we will define a characteristic function for
the occurrence of each collision outcome (accretion, rebound or fragmentation
respectively) as a function of the impact velocity. The simplest way to achieve
this is to use step functions of the square velocity dispersions $T_i$.
\hfil
\vskip .1cm
The accretion will be characterized by the function  $\Phi_A$ defined with the
help of the exhaustive rule as:

$$  \Phi_A = 1 - \Phi_F - \Phi_R  ~ ,\eqno (4.7)  $$

where $\Phi_F$ and $\Phi_R$ are the characteristic functions for the
occurrence of fragmentation or rebound (respectively) and are given by:

$$ \Phi_F(i,j) = H\left({{T_i + T_j}\over T_F}\right) \eqno
(4.8) $$

and

$$ \Phi_R(i,j)  =  H\left({{T_i + T_j}\over T_D}\right)
              H\left({{T_i+T_j} \over T_R}\right)~, \eqno
(4.9) $$

where $H$ denotes the Heaviside function which vanishes for arguments
smaller than 1.
\hfil\par

\vskip .7cm

\line{V) Mass and energy budget of the fragmentation \hfill}

\vskip .3cm
In paper I planetesimals were assumed indestructible spheres which can only
stick one another; their mass distribution was described with a finite number
of populations defined by a mass $m_i$ ($i=1,..,N$) and a velocity dispersion
$\sqrt{T_i}$. Now planetesimals are allowed to fragment but the same discrete
description of the mass spectrum and of the velocity dispersions will be kept
up. The mass evolution due to the fragmentation will be modeled under the
assumption that the mass lost by a planetesimal is redistributed into fragments
of a planetesimal size; that is, each fragment is assumed to belong to one of
the N populations defined in the previous step.
\hfil
\par
In other words, fragmentation removes mass from a given population and adds
bodies to the populations of smaller masses (in accordance with the fragment
distribution); so, in our model, planetesimal accumulation will proceed under
the competition between capture and ejection of smaller bodies.
\hfil
\vskip .1cm
Now we will examine in more details the mass and the energy budget.\hfil

\vskip .5cm

\line{1) The mass budget \hfil}

\vskip .2cm
In order to make the model still more tractable we will further assume that
catastrophic destructions are only core type, that is fragmentation leads
always to a single residual body (the "core") and to smaller fragments; in the
case of supercatastrophic collisions this is a rough assumption and the "core"
will be defined, with an abuse of the standard language, as the largest
fragment.
\hfill
\vskip .4cm

\line{a) The core \hfill}

\vskip .2cm
The mass of the residual body which remains after the ejection of the fragments
is an important parameter which determines the amount of the mass loss. The
laboratory experiments have shown that, for decimetric targets, the mass of the
core depends on the relative impact energy per unit mass through a power law
relation which writes approximately:

$$m_{core} \propto m_t \left({K\over {m_t + m_p }}\right)^{\lambda}
  \eqno(5.1)$$

\vskip .1cm
where $\lambda \simeq {1.24}$, $m_t$ is the mass of the target and $m_p$ the
mass of the projectile. For these centimeter-sized gravity-free bodies the mass
of the core is approximately proportional to  $[K/Q_D(i,j)]^{\lambda}$.
\hfil
\par
In the case of the planetesimals gravitational effects can become important
through the lithostatic stresses and the self-gravity of the fragments; then
catastrophic collisions are less destructive than for gravity-free bodies with
the formation of larger cores and the production of smaller amounts of
fragment mass. In order to account roughly for the reaccumulation effects and,
in the absence of any better assumption, we will extrapolate the above
relation to the case of the planetesimals replacing directly the disruption
threshold $Q_D$ by the fragmentation threshold $Q_F$.
\hfil
\par
Then, if the impact energy $K_{ij}$ is greater than or equal to the
fragmentation threshold $Q_F(i,j)$, the ratio of the mass of the core to the
mass of the parent planetesimal will be defined by the relation:

$$ {{m^c_{ij}}\over {m_j}} = {1\over 2} \left( {{K_{ij}}\over{Q_F(i,j)}}
\right)^{-\lambda}~. \eqno (5.2) $$

The greatest value of this ratio, when $K_{ij}$ is just equal to the
fragmentation threshold, is 1/2 as assumed in G78. Obviously for the largest
planetesimals this relation is quite speculative since, in fact, the mass of
the core depends in a complex way on ejection dynamics and reaccumulation
processes; however this weakness of the model is not a crucial one.
\hfil
\par
On the other hand, if $K_{ij}$ is smaller than $Q_F$, no fragmentation can
occur and the erosion mechanisms (as for example ejection from impact craters)
will be neglected as unimportant as regard to the evolution as a whole.
\hfil
\vskip .2cm
Then, from the mass of the core it is straightforward to obtain the mass loss
which writes:

$$ m_{j} - m^c_{ij} = m_j R_{ij} \eqno (5.3)$$

where $R_{ij}$ define a "loss matrix". \hfil
\vskip .2cm
The above expressions have been derived only for head-on collisions which are
less probable events than oblique ones. The impact experiments performed by
Fujiwara and Tsukamoto (1980) indicate that oblique impacts are less efficient
than head-on impacts for the disruption of the target. In the case of the
asteroids, Davis et al (1985) accounted for this effect and derived a mean
expression for the modified core mass; the overestimate due to the head-on
approximation is important for the very catastrophic events. In the
planetesimal case the obliquity of the impacts will be taken into account
approximately replacing, after Davis et al (1985), the mass of the core
$m^c_{ij}$ by the simple expression:

$$ 3~\left({{m^c_{ij}}\over {m_j}}\right)^{2/3}~-~2~\left({{m^c_{ij}}\over
{m_j}}\right)~.
\eqno (5.4)$$

\vskip .4truecm

\line{b) The fragments \hfill}

\vskip .2truecm
The fragment mass distribution is commonly fitted by a power law whose exponent
ranges approximately from $2/3$ for barely catastrophic collisions (with
$m_c/m_j \simeq 1/2$) to 1 for completely catastrophic collisions (with
$m_c/m_j~<<~1$). G78 found that the expression:

$$b_{jk} = \left(1 + {{m^c_{jk}}\over {m_j}}\right)^{-1} \eqno (5.5) $$

where $m_j$ is the target mass and $m_k$ the projectile mass, can reproduce
satisfactorily the exponent of this power law.
\hfil
\par
The dependence of this exponent upon $Q_F$ (through equation (5.2)) indicates
that the way in which we have accounted for self gravity modifies indirectly
the mass distribution of the fragments. It is obvious, as noticed above, that
the self gravity can distort the mass distribution of the ejecta through
selective capture but, on the other hand, it is uncertain that our rough model
can describe appropriately such effect; anyway a more detailed description of
this problem is out of the scope of the present paper.
\hfil
\par
With this power law relation the number of the fragments $m_i$ which comes from
the destruction of a planetesimal $m_j$ after an impact with another one $m_k$
will be written:

$$G_i^j(k) = g_{jk} \left({\mi \over \mo}\right)^{- b_{jk}} \eqno (5.6)$$

where the normalization parameter $g_{jk}$, derived from the conservation of
the mass before and after the impact, writes:

$$ g_{jk} = {\mj \over m_{ej}}~ R_{jk}\eqno (5.7) $$

in which:

$$ m_{ej} = \sum_{l=1}^{l_{max}} \ml~\left({\ml \over \mo}\right)^{-
b_{jk}}
\eqno (5.8) $$

is proportional to the total mass of the ejected fragments; $l_{max}$ (the
index of the largest fragments) is obtained as the maximum value of the index
for which $m_c/m_l~<~1$. The above expressions are defined only if $i<j$
(fragments are smaller than the target); on the other hand as the fragments of
the smallest planetesimals are outside our discrete mass spectrum we will set
up the relation $G_i^1(k) = 0$ for any value of $i$ and $k$ .
\hfil
\par
It will be also convenient to write the total number of the fragments $m_i$
produced in a single collision as:

$$ G_i^{jk} = G_i^j(k) + G_i^k(j)~. \eqno (5.9)$$

\vskip .5cm

\line{2) The energy budget \hfil }

\vskip .2cm
The laboratory experiments (Gault and Heitowit, 1963; Fujiwara, 1980; 1982)
provide some scarce informations on the ejecta velocity distribution; they
indicate that the core have a very low velocity whereas the smallest fragments,
which originate mainly from the surface of the target, have the largest ones;
in a more general way it seems that the larger the fragment the smaller the
ejection velocity. Such conclusions could support roughly the idea of an
equipartition of the energy (still available after the disruption) between the
various fragments, as was suggested by Wiesel (1978) and Fujiwara (1980), in
fact the experimental results of Nakamura and Fujiwara (1991) seem to show
that the velocity distribution of the fragments could be fitted by a power-law
with an index equal to -1/6. So, concerning the partition of the kinetic
energy among the ejectas, we will make the two following assumptions:
\hfil
\vskip .2cm
(i) the core have a zero relative energy (as assumed by Fujiwara (1982)),
\hfil
\vskip .1cm
(ii) the fragments share between them all the energy still available after the
disruption of the parent body proportionally to the 2/3 power of their mass.
\hfil
\vskip .2cm
Obviously the velocity distribution of the ejectas issued from a large body
depends strongly on self gravity effects; however, as mentioned above, the
account of the ejecta motion will be omitted as being an additional
complication unjustified as regard to the roughness of all the other
assumptions.
\hfil

\vskip .5cm

\line{3) The evolution of the smallest bodies \hfil}

\vskip .2cm
In the above model it is clear that a problem arises with the smallest bodies
since, strictly speaking, their fragments cannot be included into any discrete
description of the mass spectrum. In order to get round this problem we
modified somewhat our integration procedure in the following way.
\hfil
\par
For a given time step the smallest planetesimals are assumed to fragment only
into identical bodies (with a core size); then, at the beginning of the next
step, the mass and the energy of this "transient population" are reprocessed
into the new population of the smallest bodies.
\hfil
\par
It is clear that this rough description of the evolution of the smallest bodies
is one of the most severe limitation of our model, especially in the case of
the supercatastrophic collisions.
\hfil

\vskip .7cm

\line{VI) The generalized evolution equations \hfill}

\vskip .3cm
The fragmentation of the planetesimals will be included in the accumulation
scenario by addition of new collision terms to the evolution equations derived
in paper I (equations (4.7), (4.8) and (4.14) of that paper).
\hfil
\vskip .5cm

\line{1) The fragmentation collision terms \hfil}
\vskip .2cm
In some sense the fragmentation of the planetesimals is similar to a simple
creation-disparition process; the associate mass and energy variation rates
allow to derive easily the various collision terms which will be used to
complete our model.
\hfil
\vskip .1cm
$\bullet$ On the one hand fragmentation increases the number of the small
bodies at the expense of the mass of the large ones. The new collision terms
which account for this mass redistribution will be written:

$$ {\left({{d\ni}\over {dt}}\right)}_{Frag} = {1\over 2} \sum_{j,k}
\omega_{jk}
n_k f_{jk} G_i^{jk} \eqno (6.1) $$

$$ {\left({{d\mi}\over {dt}}\right)}_{Frag} = -~\mi \sum_{j} \omega_{ij}
n_j
f_{ij} R_{ij} \eqno (6.2) $$

where $\omega_{ij}= (R_i + R_j)^2{T(0)}^{1/2}$ is the geometrical collision
rate which accounts for the evolution of the cloud thickness and $f_{ij} = (1 +
\Theta_{ij})\Phi_F(i,j)$ is a fragmentation coefficient in which the
$\Theta_{ij}$ are the well known Safronov parameters.
\hfil
\par
The first term (6.1) corresponds to a creation rate; that is to a total number
of fragments, with mass $m_i$, issued per second from all possible pairs of
colliding planetesimals. The second term (6.2) corresponds to the mass loss of
the ith population which results from the catastrophic collisions between a
target $m_i$ and any smaller projectile.
\hfil
\vskip .3cm

$\bullet$ On the other hand fragmentation also contributes to the evolution of
the planetesimal kinetic energy. The energy variation rate of a given
population has been derived from a simple balance between the energy supplied
by fragments $m_i$ issued from larger bodies and the energy loss which results
from the disparition of the destructed bodies; we obtained:

$$ \left({{dK_i}\over{dt}}\right)_{Frag} = {1\over 2}\sum_{jk} \omega_{jk}
n_j
n_k \epsilon_{jk} f_{jk} G_i^{jk}
- {1\over 2} n_i m_i T_i \sum_j \omega_{ij} n_j f_{ij}~{{m_{ij}^*}\over
{m_i}}
\eqno (6.3) $$

where:

$$\epsilon_{jk} = {\chi\over {\Sigma^F_{jk}}} {1\over 2} m_{jk}^* (T_j +
T_k) \left({m_i}\over{m_o}\right)^{2\over3} \eqno (6.4) $$

is the kinetic energy each fragment carries away after the partition (in
accordance with the velocity distribution of the fragments) of the energy still
available after the disruption and with also:
\hfil

$$   \Sigma^F_{jk} = \sum_i G_i^{jk} \left({m_i\over
m_o}\right)^{2\over 3} ~. \eqno (6.5) $$

\vskip .5cm

\line {2) The complete set of the equations \hfil}

\vskip .2cm

When both accretion and fragmentation operate the evolution equations for the
mass and the energy write:

$$ {{d\ni}\over {dt}} = - \sum_j n_i n_j \omega_{ij} a_{ij} C_{ij} +
       {1\over 2} \sum_{j,k} n_k \omega_{jk} f_{jk} G_i^{jk} \eqno (6.6)$$

$$ {{d\mi}\over {dt}} = \sum_j \omega_{ij} n_j (m_j a_{ij} A_{ij} -
m_i f_{ij} R_{ij})  \eqno (6.7)$$

$$ {{dK_i}\over{dt}}  = \left({{dK_i}\over{dt}}\right)_{Acc} +
\left({{dK_i}\over{dt}}\right)_{Frag} +
\left({{dK_i}\over{dt}}\right)_{Enc}
 \eqno (6.8) $$

in which the contributions of the accretion and of the encounters, reported
from paper I, are given by:

$$ \left({{dK_i}\over{dt}}\right)_{Acc} =
{{m_i}\over{2\sqrt{2}}}\sum_{j=1}^N
A_{ij}{{\omega_{ij}  n_j m_j }\over{(m_i + m_j)^2}} \left[(m_j T_j - m_i
T_i) -  (m_i + m_j) T_i \right] \I ~\Phi_A(i,j) ~~~~ \eqno (6.9) $$

and

$$ \left({{dK_i}\over{dt}}\right)_{Enc} = {2\sqrt{2}}~ m_i \sum_{j=1}^N
{{\omega_{ij} n_j\mj}\over{(\mi +\mj)^2}} \Theta ^2_{ij} L_{ij}
\bigl[\left(\mj\Tj - \mi\Ti\right)\H + {1\over 2}\mj(\Ti~+ \Tj) \J \bigr]
 \eqno (6.10) $$

in the above equations we used the following matricial notations:

$$ A_{ij} = \cases {1, &if $j\le i$~;\cr 0, &otherwise.\cr}$$

for the mass accretion, \hfil

$$ C_{ij} = \cases {1, &if $j\ge i$~;\cr 0, &otherwise.\cr}$$

for the capture of bodies, \hfil
\par
and we defined also the accretion coefficient:  $a_{ij} = (1 + \Theta_{ij})
\Phi_A (i,j)$. \hfil

\vskip .7cm

\line{VI) Planetesimal growth with allowance for fragmentation \hfill}

\vskip .4cm
In paper I colliding planetesimals were only allowed to stick one another and
integration was stopped when the increasing velocity dispersions reached
$500~kms^{-1}$, a value assumed large enough for fragmentation to play a role.
\hfil
\vskip .1cm
Then, in section II, the new accretion calculations were carried out (see
figures 1, 2 and 3) until a single body remains in the largest batch; below,
the results of these calculations, in which no fragmentation comes into play,
will be used to drive meaningful comparisons.
\hfil
\par
Now each collision outcome have an occurrence (defined in section IV) which
changes as a function of time according to the evolution of the velocity
dispersions. The integration of the generalized evolution equations has been
performed with the same procedure as in paper I (except the above modifications
necessary to account for the fragmentation of the smallest bodies) and with a
resolution on the mass distribution given by 12 different batches. In the
course of the computations, conservation of mass and energy (kinetic energy
partitioned between the various ejecta) were controlled at regular time
intervals.
\hfil
\vskip .2cm
In order to proceed with increasing complexity and to make appropriate
comparisons with the results obtained for indestructible bodies, we will
distinguish between two cases according to the elasticity of the collisions.
\hfil
\par
\vskip .5cm

\line{1) The case of completely inelastic collisions \hfil}

\vskip .2cm
The relaxation of the mass spectrum, set out in section II, reduces the range
of initial mass distributions to very steep functions. However initial
conditions differ also by the choice of the characteristic size of the
primordial planetesimals; these bodies which resulted from the agglomeration of
the dust particles in the first age of the protoplanetary nebula (a
sedimentation stage followed by a gravitational instability after the
traditional Safronov-Goldreich and Ward's scenario) could have an average size
of the order of some kilometers at the earth distance. In order to discuss the
importance of this initial size, it will be convenient to distinguish between
two different cases, each one corresponding to a well defined characteristic
size: 1 km or 10 km.
\hfil
\vskip .1cm
In the beginning, the velocity dispersions are too small for fragmentation to
operate and the evolution of the planetesimals looks like that of the
indestructible bodies investigated above (this appears clearly when figures 1,
2 and 3 are compared with the figures 6, 7 and 8). Then, as the velocity
dispersions increase as a function of time, fragmentation comes gradually into
play and modifies the course of the evolution.
\hfil
\vskip .2cm

$\bullet$ Primordial planetesimals with a characteristic size of the order
of $10~km$ \hfil
\vskip .1cm
In the size range of the Safronov exponential mass spectrum, two cases have
been distinguished following the disruption energy $Q_D$ is assumed dependent
or independent of the impact velocity.
\hfil
\par
In both cases calculations were carried out until a single body (with a size of
the order of 1000km) remains in the largest batch. Fragmentation is found to
play a role only at the end, in relation with the large values of the velocity
dispersions. The number of bodies and the velocity dispersions obtained in
these two cases are plotted on figures 6 when the disruption energy is assumed
independent of the impact velocity and on figures 7 when the opposite
assumption holds.
\hfil
\par
Fragmentation comes into play after a rapid growth stage (the size of the
largest bodies increase by a factor 100 in a time-scale of the order of
$3.10^5~yrs$) and modifies the evolution of the mass spectrum in the following
way:
\hfil
\vskip .1cm
- a tail comes at the small size end; it contains a fraction of the total mass
which ranges from $7\%$ to $17\%$, following $Q_D$ is independent or dependent
of the impact velocity, respectively; at the end, the tail evolves faster and
faster due to the increasing importance of the destruction processes.
\hfil
\vskip .1cm
- the growth of the largest bodies slows down but a single embryo succeeds in
accreting $17\%$ of the mass of the swarm; the growth time scale is of the
order of $3.7~10^5~yrs$ or $6.2~10^7~yrs$, following $Q_D$ is independent or
dependent of the impact velocity, respectively; the growth is slower and slower
as fragmentation becomes more and more efficient.
\hfil
\vskip .2cm
On the other hand the evolution of the velocity dispersions is only weakly
modified as compared to the accretion of indestructible planetesimals: the
small bodies of the tail have velocity dispersions joining up smoothly with the
rest of the spectrum and the largest ones can reach the Hill velocity.
Nevertheless the final values of the velocity dispersions are slightly smaller
than in the case of indestructible bodies.
\hfil
\vskip .2cm

$\bullet$ Primordial planetesimals with a characteristic size of the order
of $1~km$ \hfil
\vskip .1cm
In the size range of the power law mass distributions defined in section II,
planetesimals are more brittle bodies and only the case of a disruption energy
independent of the impact velocity has been investigated.
\hfil
\par
Fragmentation is found to modify the evolution of the mass distribution much
more strongly than in the 10 km size range; this is obviously due to the fact
that one-kilometer-sized planetesimals are the most brittle bodies
(this appears clearly on Fig. 4). The calculations were
performed in the same way as above; the resulting number of bodies and velocity
dispersions are plotted on figures 8 as a function of mass and for various time
steps. The evolution have approximately the same trend as in the ten-kilometer
size range but fragmentation takes place earlier. The rapid growth stage, in
which the size of the largest bodies increases by a factor 100, lasts only some
$10^4~yrs$ but on the other hand, the stage in which fragmentation comes into
play is much more longer. After $4.10^7~yrs$, the largest batch contains $12\%$
of the total mass, that is approximately 16 bodies with a size of the order of
$670~km$, and the tail only some $7\%$ . A great amount of the total mass is
found in the intermediate scales.
\hfil
\par
As expected from the previous case, if the disruption energy is assumed
dependent on the impact energy the planetesimals are still more brittle and the
importance of the catastrophic collisions is still stronger.
\hfil

\vskip .5cm

\line{2) The case of weakly elastic collisions \hfil }

\vskip .2cm
The inelasticity of the planetesimal collision is likely very large due to the
surface composition (presence of a regolith layer) but also to the internal
structure of these bodies. However, in order to test the importance of this
parameter, we have set up the restitution coefficient $\gamma_o$ equal to 0.2,
a value in the range of the largest values assumed by G78.
\hfil
\vskip .1cm
A priori considerations suggest that rebounds could have important consequences
on the evolution of the planetesimals and figures 2 testify also that rebounds
can operate in some range of the impact velocity. For example, rebounds are
expected to modify the evolution of the mass spectrum at the small size end
because they could hinder the growth of the small bodies and their own capture
by the larger ones; so, the mass distribution should be expected steeper than
for completely inelastic collisions. It is also to be noted that fragmentation,
which results in the formation of small body populations, could strengthen the
importance of the rebounds. In fact computations performed in the same way as
above seem to show that, in the investigated size range, the inelasticity of
the collisions play a negligible role.
\hfil

\vskip .7cm

\line{VIII) Discussion and conclusions \hfil }

\vskip .4cm
This work improves and completes the model of the planetesimal accumulation
started in paper I. Mass dispersion among the largest bodies takes an important
part in the evolution of initially gentle mass distributions. For example,
simple power laws are found to relax rapidly into a very steep function. Such a
relaxation of the mass distribution comes from the kinetics of the coagulation
process and is similar, in some respect, to the relaxation of the distribution
of the velocity dispersions which is driven by the dynamical friction (see
paper I).
\hfil
\par
\vskip .1cm
This transient behaviour of the mass and velocity distributions, whose detailed
study is out of the scope of the present work, shows that the first stage of
the evolution of the planetesimals is nearly independent of the initial
conditions; the swarm itself supplies us rapidly with appropriate initial
conditions. Afterwards the evolution of the planetesimals proceeds
approximately in the same way as in paper I when we started from an
exponential mass distribution: a rapid growth of the largest bodies with a
simultaneous increase of the velocity dispersions (such behaviour occurs
because gradually gravitational stirring can overcome dynamical friction).
This increase of the velocity dispersions strongly suggests that fragmentation
have to play a role in the accumulation scenario of the planets.
\hfil
\par
\vskip .3cm
When catastrophic destructions are possible outcomes of the two-body
planetesimal collisions, the evolution of the mass distribution proceeds, in
qualitative agreement with the work of Beaug\'e and Aarseth (1990), into two
different stages. During the first one accretion is the dominant process and
results in the rapid growth of much larger bodies; during the second one
fragmentation counteracts the effect of the accretion and the evolution of the
mass distribution have two different aspects:
\hfil
\par
- at the small scales the mass spectrum stretch itself with the formation of a
tail, a behaviour in agreement with the results of G78 and of WS89 (see Figs.
6);
\hfil\par
- at the large scales "brittle" planetesimals grow on a longer time scale than
indestructible ones; this new aspect of the accumulation is consistent with the
conclusion of Beaug\'e and Aarseth (1990) according to which the formation of
new embryos stalls when fragmentation begin to operate.
\hfil
\par
\vskip .1cm
Both the extent of the tail and the growth time-scale are found to depend on
the fragmentation properties of the planetesimals.
\hfil
\par
\vskip .1cm
Obviously during this second stage, the velocity dispersions evolve in relation
with the changes of the mass spectrum into two different ways:
\hfil
\par
- accretion tends to transfer most of the total mass from the small scales to
the larger ones and, so, keeps going the gravitational stirring;
\hfil
\par
- fragmentation tends to transfer most of the total cross-section of the
inelastic collisions to the smaller scales and, thus, controls the cooling of
the swarm.
\hfil
\par
The velocity dispersions evolve more slowly as when fragmentation is not
allowed but their distribution among the various populations remains
approximately the same.
\hfil
\par\vskip .1cm
It results from this two fold evolution that the collisions with the small
bodies, although the most probable ones, cannot contribute significantly to
the growth of the planetesimals as the small body populations contain only a
small amount of the total mass. On the other hand, due to the increasing
velocity dispersions, fragmentation is at work up to a maximum size which
increases as a function of time; when this size reaches the intermediate
scales of the mass spectrum a significant amount of the total mass is affected
and streams much less rapidly to the large scales. These two reasons explain
why planetesimals grow more slowly when fragmentation comes at work.
\hfil
\par
\vskip .1cm
The second stage of the evolution also depends on the characteristic size of
the primordial planetesimals:
\hfil
\par
\vskip .1cm
- when set up of the order of 10 km, fragmentation cannot prevent the growth of
a single planetary embryo (with some 17\% of the total mass and a radius of the
order of $1000~km$) which remains embedded in a swarm of much smaller bodies,
\hfil
\vskip .1cm
- when set up of the order of 1 km, fragmentation is easier and operates
earlier than for 10-kilometer sized bodies but the evolution is much more
slowly.
\hfil
\par
\vskip .2cm
As a result catastrophic collisions tend to slow down significantly the growth
of planetary embryos. The growth time-scales depend both on the initial size of
the primordial planetesimals and on their effective strength all along their
growth.
\hfil
\vskip .1cm
Following the classical formation model (Safronov, 1969; Goldreich and Ward,
1973), the characteristic size of these primordial bodies is of the order of
some kilometer at the earth distance but increases with increasing distances
from the sun (some hundred kilometers at the Uranus distance); if so,
catastrophic collisions could slow down the growth of the embryos much more
effectively in the inner parts of the solar system than in its outer parts. It
must be pointed out that such conclusion is consistent with the
accumulation scenario of the giant planet cores and of the outer planets.
\hfil
\vskip .1cm

Following the recent fragmentation model of Housen et al (1991), the effective
strength of asteroidal bodies is likely velocity dependent. As appears on
figures 6 and 7 such a dependence have a significant consequence on the
accretion time-scale (it is of the order of $10^5~yrs$ or $10^7~yrs$ following
the strength is velocity dependent or not, respectively). However, as regard to
the crudeness of the present model, this slowing down of the
growth of the planetary embryos does not seems strong enough for the time-scale
problem of the formation of the outer planets to be reopened. Nevertheless this
could be the case if the characteristic size of the primordial
planetesimals was smaller than one kilometer; then the formation of the
planetesimals would play a much more important role than previously believed
since the size of the bodies build during this stage could
control the growth time-scale of the planetary embryos and, so, the fate of
protoplanetary clouds evolution.
\hfil
\vskip .1cm
To summarize, in order to form a planetary system, the growth of massive
embryos should take place before catastrophic collisions become too effective;
that is accretion should bring most of the total mass to the largest
scales before fragmentation have time to shift most of the total cross-section
to the small scales.
\hfil
\par
At the end of our integration some large bodies are found to drop out of the
mass distribution; this fact seems consistent with the idea (Safronov,
1966; Lissauer and Safronov, 1991) that the largest bodies to be captured by
the proto-earth should have masses of the order of $10^{-3}$ or $10^{-2}$ earth
masses as to explain planetary spins.
\hfil
\par
\vskip .25cm
Now, we will also mention the implicit assumptions made in the above
calculations:
\hfil
\par
- a 3-body approximation of the accretion rate was omitted since in paper I
such a refinement was found unimportant in the overall evolution of  the
planetesimals.
\hfil
\vskip .1cm
- integration was pursued till a single body remains in the population of the
largest bodies; in fact it is clear that a problem arises in such a situation
since present theory is inappropriate to describe the change in the velocity
dispersions of the small bodies. Nevertheless the work of Lecar and Aarseth
(1986) supports the idea that, at the end of the evolution, long range
perturbations between planetary embryos allow secular increase of the velocity
dispersion and, so, can avoid premature isolation.
\hfil

\vskip .2cm
Finally let us make two comments about the formalism used in this paper:
\hfil
\par
\vskip .1cm
- The basic statistical assumptions are well suited in the beginning of the
evolution (since the initial number of bodies is very large) but become poorer
and poorer at the end, when the population of the largest bodies reduces
strongly; then (as pointed out by Safronov, 1969), if the mass of the embryos
is of the order of the mass of all other bodies inside a given zone, strong
perturbations of the small body orbital motions are expected and the formalism
must be changed.
\hfil
\vskip .1cm
- The accumulation process neglects the spatial fluctuations in the number
density of the planetesimals; in this homogeneous mean-field description the
accretion calculations are easier but the planetary embryos can form anywhere
in the system.
\hfil
\par
\vskip .1cm
The next step we have begun to investigate is the study of non-homogeneous
accumulation processes and angular momentum transfers, two underlying questions
in the final stage of the accumulation scenario.
\hfil
\vskip .1cm
Obviously other constraints for planet formation are also to be found in the
present high velocity dispersions of the asteroid belt, likely in connection
with the date of birth of a massive Jupiter core.
\hfil\par

\vskip 1.cm

- Acknowledgments: \hfil\par
\vskip .2cm
This work was supported by I.N.S.U. under grant   A.T.P. 674 AP 88 87.67.19 and
calculations were performed on the computer  CYBER 2000 of C.N.E.S. .
Constructive criticisms of the manuscript were provided by P. Paolicchi and R.
Greenberg. One of the authors  (P.B.) thanks R. Greenberg for helpful
discussions.
\hfil
\par

\vfil
\eject

\vskip .5cm
\line {REFERENCES \hfill}
\vskip .5cm
 . ANDERS, E. (1965). Fragmentation History of Asteroids. Icarus 4, 399- 409.
\hfill\par
\vskip .1cm
 . BARGE, P., and R. PELLAT (1990). Self consistent velocity dispersions and
mass spectrum in the protoplanetary cloud.  Icarus 85, 481-498.
\hfill\par
\vskip .1cm
 . BARGE, P., and R. PELLAT (1991). Mass Spectrum and Velocity Dispersions
during Planetesimals Accumulation (I - Accretion). Icarus 93, 270-287.
\hfill\par
\vskip .1cm
 . BEAUG\'E C. and S.J. AARSETH (1990). N-Body Simulations of Planetary
Formation. Mon. Not. R. Astron. Soc. 245, 30-39.
\hfill\par
\vskip .1cm
 . BIANCHI, R., F. CAPACCIONI, P. CERRONI, M. CORADINI, E. FLAMINI, P. HURREN,
G. MARTELLI and P.N. SMITH (1984). Experimental Simulation of Asteroidal
Fragmentation by Macroscopic Hypervelocity Impacts against Free Falling Bodies.
Astron. Astrophys. 139, 1-6.
\hfill\par
\vskip .1cm
 . CHAPMAN, C.R. and D.R. DAVIS (1975). Asteroid Collisional Evolution:
Evidence for a much Larger Population. Science (Washington, D.C.) 190, 553-556.
\hfill\par
\vskip .1cm
 . DAVIS, D.R., and C.R. CHAPMAN (1977). The collisional evolution of asteroid
compositional classes. Lunar Planet. Sci. VIII, 224.
\hfill\par
\vskip .1cm
 . DAVIS, D.R., CHAPMAN, C.R., GREENBERG R. and S.J. WEIDENSCHILLING (1979).
Collisional evolution of asteroids: Populations, Rotations and velocities. In
Asteroids (T. Gehrels, Ed.), PP 528-557. Univ. of Arizona Press, Tucson.
\hfill\par
\vskip .1cm
 . DAVIS, D.R., CHAPMAN, C.R., WEIDENSCHILLING, S.J. and R. GREENBERG (1985).
Collisional History of Asteroids: Evidence from Vesta and the Hirayama
Families. Icarus 62, 30-53.
\hfill\par
\vskip .1cm
 . DOHNANYI, J. S. (1969). Collisional model of asteroids and their debris. J.
Geophys. Res., 74, 2531-2554.
\hfill\par
\vskip .1cm
 . FARINELLA, P., P. PAOLICCHI and V. ZAPPALA (1982). The Asteroids as Outcomes
of Catastrophic Collisions. Icarus 52, 409-433.
\hfill\par
\vskip .1cm
 . FARINELLA, P., A. MILANI, A. NOBILI, P. PAOLICCHI AND V. ZAPPALA (1983).
Hyperion: Collisional Disruption of a Resonant Satellite. Icarus 54, 353- 360.
\hfill\par
\vskip .1cm
 . FUJIWARA, A., G. KAMIMOTO, and A. TSUKAMOTO (1977). Destruction of basaltic
bodies by high velocity impact. Icarus 31, 277-288.
\hfill\par
\vskip .1cm
 . FUJIWARA, A. (1980). On the mechanism of catastrophic destruction of minor
planets by high-velocity impact. Icarus 41, 356.
\hfill\par
\vskip .1cm
 . FUJIWARA, A. and  A. TSUKAMOTO (1980). Experimental Study on the Velocity of
Fragments in Collisional Breakup. Icarus 44, 142.
\hfill\par
\vskip .1cm
 . FUJIWARA, A. (1982). Complete fragmentation of the parent bodies of Themis,
Eos, and Koronis Families. Icarus 52, 434.
\hfill\par
\vskip .1cm
 . GAULT, D., E., and E. D. HEITOWIT (1963). The partition of energy for
hypervelocity impact crater, formed in rocks. In Proceedings of the Sixth
Hypervelocity Impact Symposium, Vol 2, pp 419-456. Firestone Rubber Co.,
Cleveland, Ohio.
\hfill\par
\vskip .1cm
 . GAULT, D. E., and J., A. WEDEKIND (1969). The destruction of tektites by
micrometeoroid Impact. J. Geophys. Res. 74, 6780-6794.
\hfill\par
\vskip .1cm
 . GRADIE, J.C., and B. ZELLNER (1977). Asteroids families: Observational
evidence for common origins. Science 197, 254-255.
\hfill\par
\vskip .1cm
 . GREENBERG, R., J. F. WACKER, W. K. HARTMANN, and C. R. CHAPMAN (1978)
Planetesimals to planets: Numerical simulation of collisional evolution. Icarus
35, 1-26.
\hfill\par
\vskip .1cm
 . HELLYER, B. (1970). The Fragmentation of the Asteroids. Monthly Notices R.
Astron. Soc. 148, 383.
\hfill\par
\vskip .1cm
 . HARTMANN, W.K. (1978). Planet Formation: Mechanism of Early Growth. Icarus
33, 50-61.
\hfill\par
\vskip .1cm
 . HORNUNG, P., R. PELLAT, and P. BARGE (1985)  Thermal velocities equilibrium
in the protoplanetary cloud. Icarus 64, 295-307.
\hfill\par
\vskip .1cm
 . HOUSEN, K.R. and K.A. HOLSAPPLE (1990). On the Fragmentation of Asteroids
and Planetary Satellites. Icarus 84, 226-253.
\hfill\par
\vskip .1cm
 . HOUSEN, K.R., R.M. SCHMIDT, and K.R. HOLSAPPLE (1991). Laboratory
Simulations of Large Scales Fragmentation Events. Icarus 94, 180-190.
\hfill\par
\vskip .1cm
 . IP, W.H. (1979). On Three Types of Fragmentation Processes Observed in the
Asteroid Belt. Icarus 40, 418-422.
\hfill\par
\vskip .1cm
 . LECAR M. and S.J. AARSETH (1986). A Numerical Simulation of the Formation of
the Terrestrial Planets. Astrophys. J. 305, 564.
\hfill\par
\vskip .1cm
 . NAKAMURA, A. and A. FUJIWARA (1991). Velocity Distribution of Fragments
Formed in a Simulated Collisional Disruption. Icarus 92, 132-146.
\hfill\par
\vskip .1cm
 . PECHERNIKOVA, G.V. (1975). Mass Distribution of Protoplanetary Bodies. I.
Initial datas for numerical solution. Sov. Astron., Vol. 18, No. 6, 778- 783.
\hfill\par
\vskip .1cm
 . PECHERNIKOVA, G. V., V. S. SAFRONOV, and E. V. ZVYAGINA (1976). Mass
distribution of protoplanetary bodies. II. Numerical solution of generalized
coagulation equation. Sov. Astron., Vol. 20, No. 3, 346-350.
\hfill\par
\vskip .1cm
 . PIOTROWSKI, S. (1953). The collisions of asteroids. Acta. Astron., Ser. a.
Vol. 5, 115-138.
\hfill\par
\vskip .1cm
 . SAFRONOV, V.S. (1969). Evolution of the Protoplanetary Cloud and Formation
of the Earth and the Planets. Israel program for scientific translation, TT-F
677.
\hfill\par
\vskip .1cm
 . SAFRONOV, V.S. (1966). Sizes of the largest bodies falling onto the planets
during their formation. Sov. Astron., Vol. 9, No. 6, 987-991.
\hfill\par
\vskip .1cm
 . WETHERILL, G. W., and G. R. STEWART (1989) Accumulation of a swarm of small
planetesimals. Icarus 77, 330-357.
\hfill\par
\vskip .1cm
 . WIESEL, W. (1978). Fragmentation of Asteroids and Artificial Satellites in
Orbit. Icarus 34, 99-116.
\hfill\par
\vskip .1cm
 . ZAPPALA, V., FARINELLA, P., KNEZEVIC, Z. and P. PAOLICCHI (1984).
Collisional Origin of the Asteroid Families: Mass and Velocity Distributions.
Icarus 59, 261-285.
\hfill\par
\vskip .1cm
 . ZVYAGINA, E. V., G. V. PECHERNIKOVA, and V. S. SAFRONOV (1974). Qualitative
solution of the coagulation with allowance for fragmentation. Sov. Astron. Vol.
17, No. 6, pp 793-800.
\hfil
\vfill
\eject

\topskip=.5cm
\centerline{FIGURES CAPTIONS}

\vskip .6cm

 . Figs.1 : Cumulative number of bodies and velocity dispersions as a function
of mass for different time intervals ($\theta~=~t/t_o$) starting from a
Safronov exponential mass spectrum. The collisions are purely inelastic, the
same gas drag as in WS89 is assumed. The characteristic parameters have the
following values: $m_o~=~1.15~10^{16}~kg$, $V_o~=~15.5~ms^{1}$,
$N_o~=~5.51~10^7$ and $t_o~=~1.22~10^5~yrs$ (obtained from a surface density
$\sigma~=~224~kgm^{- 2}$).
\hfil
\vskip .3cm
     - Fig.1a : Evolution of the cumulative number of bodies
\hfil
\vskip .2cm
     - Fig.1b : Evolution of the velocity dispersions \hfil
\vskip .35cm

 . Figs.2 :  Cumulative Number of bodies and velocity dispersions as a function
of mass for different time intervals ($\theta~=~t/t_o$) if the mass is
initially in the large bodies (the initial mass distribution is a power law
with an index $a~=~0.$). The collisions are purely inelastic and the same gas
drag as in WS89 is assumed. The characteristic parameters have the following
values: $m_o~= 1.6~10^{13}~kg$ (corresponding to a radius of $1~km$),
$V_o~=~T_o^{1/2} ~=~1.73~ms^{-1}$, $N_o~=~4.~10^{10}$ and
$t_o~=~1.36~10^4~yrs$.
\hfil
\vskip .3cm
     - Fig.2a : Evolution of the cumulative number of bodies
\hfil
\vskip .2cm
     - Fig.2b : Evolution of the velocity dispersions \hfil
\vskip .35cm

 . Figs.3 :  Cumulative Number of bodies and velocity dispersions as a function
of mass for different time intervals ($\theta~=~t/t_o$) if the mass is
initially in the small bodies (the initial mass distribution is a power law
with an index $a~=~-2.$). The collisions are purely inelastic, the same gas
drag as in WS89 is assumed. The characteristic parameters have the following
values: $m_o~= 1.6~10^{13}~kg$ (corresponding to a   radius of $1~km$),
$V_o~=~T_o^{1/2} ~=~1.73~ms^{-1}$, $N_o~=~4.~10^{10}$ and
$t_o~=~1.36~10^4~yrs$.
\hfil
\vskip .3cm
     - Fig.3a : Evolution of the cumulative number of bodies
\hfil
\vskip .2cm
     - Fig.3b : Evolution of the velocity dispersions \hfil
\vskip .35cm

 . Fig.4 : Specific energies required to disrupt ($Q_D^*$) or to fragment
($Q_D^*$) bodies. We have assumed an impact velocity of $5.kms^{-1}$ and an
energy transfer rate $\chi~=~0.1$ . In the right part of curve $Q_D^*$ the
target cannot be disrupted and in between curve $Q_D^*$ and curve $Q_F^*$ the
target is disrupted but the ejection of the fragments is impossible.
\vskip .35cm

 . Figs.5 : The velocity thresholds $T_R^{1/2}$, $T_D^{1/2}$, $T_F^{1/2}$ for
the occurrence of accretion, rebound or fragmentation, respectively, as a
function of the projectile to target radius ratio (see text).
\hfil
\vskip .3cm
     - Fig.5a : Case of the small bodies. Rebounds are very effective whereas
reaccumulation is inexistent.
\hfil
\vskip .2cm
    - Fig.5b : Case of the intermediate bodies. In the small strip between the
two decreasing curves reaccumulation could began to play a role. Accretion
corresponds either to a sticking or to a reaccumulation of the two bodies.
\hfil
\vskip .2cm
  - Fig.5c : Case of the large bodies. In the region between the two decreasing
curves reaccumulation is likely to play a role whereas rebounds are much less
effective.
\hfil
\vskip .35cm

 . Figs.6 : Cumulative number of bodies and velocity dispersions as a function
of mass for different time intervals ($\theta~=~t/t_o$) starting from a
Safronov exponential mass spectrum. Both accretion and fragmentation are
allowed; the disruption energy is assumed independent of the impact velocity;
the same gas drag as in WS89 is assumed. The characteristic parameters have
following values: $m_o~=~1.15~10^{16}~kg$, $V_o~=~15.5~ms^{-1}$,
$N_o~=~5.51~10^7$ and $t_o~=~1.22~10^5~yrs$ (obtained from a surface density
$\sigma~=~224~kgm^{- 2}$).
\hfil
\vskip .3cm
     - Fig.6a : Evolution of the cumulative number of bodies
\hfil
\vskip .2cm
     - Fig.6b : Evolution of the velocity dispersions \hfil
\vskip .35cm

 . Figs.7 : Cumulative number of bodies and velocity dispersions as a function
of mass for different time intervals ($\theta~=~t/t_o$) starting from a
Safronov exponential mass spectrum. Both accretion and fragmentation are
allowed; the disruption energy is assumed to depend on the impact velocity; the
same gas drag as in WS89 is assumed. The characteristic parameters have
following values: $m_o~=~1.15~10^{16}~kg$, $V_o~=~15.5~ms^{-1}$,
$N_o~=~5.51~10^7$ and $t_o~=~1.22~10^5~yrs$ (obtained from a surface density
$\sigma~=~224~kgm^{- 2}$).
\hfil
\vskip .3cm
     - Fig.7a : Evolution of the cumulative number of bodies
\hfil
\vskip .2cm
     - Fig.7b : Evolution of the velocity dispersions \hfil
\vskip .35cm

 . Figs.8 :  Cumulative Number of bodies and velocity dispersions as a function
of mass for different time intervals ($\theta~=~t/t_o$) if the mass is
initially in the small bodies (the initial mass distribution is a power law
with an index $a~=~-2.$). Both accretion and fragmentation are allowed; the
disruption energy is assumed independent of the impact velocity; the same gas
drag as in WS89 is assumed. The characteristic parameters have the following
values: $m_o~= 1.6~10^{13}~kg$ (corresponding to a radius of $1~km$),
$V_o~=~T_o^{1/2} ~=~1.73~ms^{-1}$, $N_o~=~4.~10^{10}$ and
$t_o~=~1.36~10^4~yrs$.
\hfil
\vskip .3cm
     - Fig.8a : Evolution of the cumulative number of bodies
\hfil
\vskip .2cm
     - Fig.8b : Evolution of the velocity dispersions \hfil
\vskip .35cm

\vfill
\supereject\end